# Diagonal Metrics of Static, Spherically Symmetric Fields
# The Geodesic Equations and the Mass-Energy Relation from the Coordinate Perspective

Franz-Günter Winkler

*Stumpergasse 63/11, A-1060 Vienna, Austria (Home Address)*
Phone: +43 1 5969798
Email: fg.winkler@aon.at

**Abstract** The geodesic equations for the general case of diagonal metrics of static, spherically symmetric fields are calculated. The elimination of the proper time variable gives the motion equations for test particles with respect to coordinate time and an account of "gravitational acceleration from the coordinate perspective". The results are applied to the Schwarzschild metric and to the so-called exponential metric. In an attempt to add an account of "gravitational force from the coordinate perspective", the special relativistic mass-energy relation is generalized to diagonal metrics involving location dependent and possibly anisotropic light speeds. This move requires a distinction between two aspects of the mass of a test particle (parallel and perpendicular to the field). The obtained force expressions do not reveal "gravitational repulsion" for the Schwarzschild metric and for the exponential metric.

***Keywords*** *pseudo-Riemannian geometry   gravitational repulsion   Schwarzschild metric   exponential metric   scalar gravitation*

# 1 Introduction

In his analysis of the Schwarzschild solution Hilbert [1] calculated the acceleration of a radially moving particle with respect to coordinate time. He found that – depending on speed and location – acceleration may change sign and thus point away from the center of gravitation.[1] This phenomenon has later been labeled "gravitational repulsion" and has created some controversy. An overview and a good portion of clarification has been given by McGruder [3] who lifted the limitation to mere radial motion and gave a full account of radial acceleration with respect to coordinate time.

---

[1] Simultaneously with Hilbert, Bauer [2] gave an independent first account of gravitational repulsion.

In this article we generalize these considerations on gravitational acceleration with respect to coordinate space and coordinate time to the class of all diagonal metrics of static, spherically symmetric fields. Going one step further, we extend the "coordinate perspective" by deriving expressions for gravitational force. The basis for this endeavor is a suggested adaptation of the special relativistic mass-energy relation to location dependent, anisotropic light speeds.

In the first part the geodesic equations for the class of metrics under investigation are calculated; the elimination of the proper time variable leads to the motion equations with respect to coordinate time. The results are applied to the Schwarzschild metric and to the exponential metric.[2]

In the second part we introduce concepts of energy and mass of test particles as they are seen from the coordinate perspective. The general case requires us to specify two aspects of mass (parallel and perpendicular to the field), but still gives conservation laws for energy and angular momentum and a straight-forward treatment of gravitational force.

There are two concrete result of the present analysis: (1) It turns out that the motion equations of a "scalar gravity model" (henceforth called "SG-model") by Jan Broekaert[3] [12] which is not based in the framework of pseudo-Riemannian geometry are just the geodesic equations of the exponential metric reformulated with respect to coordinate time. (2) Gravitational repulsion does not imply that the gravitational force may be repulsive.

---

[2] The exponential metric appears in the context of different alternative approaches to gravitation (e.g. Yilmaz [4], Rosen [5], Ni [6]). When judging the compliance of such theories with the experimental evidence of general relativity two statements seem to be consensual: (a) The four classical tests (gravitational redshift, light deflection, perihelion precession, and radar echo delay) are, indeed, confirmed, as long as the gravitational field is regarded to be "at rest". (b) Most of these theories postulate a "preferred frame", relative to which the solar system should be moving; for a number of such models, it has been shown [7] that this leads to different predictions which are supposedly ruled out by experiment. This conjecture is not shared by Arminjon [8], though, whose model also results in the exponential metric in the resting case.

In section 3 we will give an argument that supports statement (a). A candidate theory which might overcome the problems connected with (b) is Yilmaz's later approach [9], which claims to obey the Einstein field equations; it is being regarded as highly controversial, though (see the dispute with Misner [10, 11]).

[3] Broekaert's model is built upon work by Sjödin and Podlaha [13, 14, 15].

# 2 Metric Scheme and Geodesic Analysis

## 2.1 Coordinate Systems and Vectors

In order to prevent misunderstandings right at the beginning, some remarks have to be made concerning the meaning of the spherical coordinates that are being used throughout the article. The calculations are applicable to metrics belonging to different theories which give different meanings to coordinates. While for some alternative approaches to gravitation $r$, $\theta$ and $\varphi$ are ordinary spherical coordinates in a 3D Cartesian coordinate system, it is well known that e.g. the Schwarzschild coordinates (i.e. the coordinates in which the Schwarzschild metric is usually expressed) only match this intuitive understanding of coordinates for the case of weak fields and for large distances from the center of gravitation. The present calculations do not depend on this question, though.

The suggested treatment of vectors for speed, acceleration, momentum and force "from the coordinate perspective" only assumes that parallel and perpendicular unit vectors $\mathbf{e}_\parallel$ and $\mathbf{e}_\perp$ can be introduced (we restrict ourselves to the ($r, \varphi$)-plane where $\theta = \pi/2$ and $d\theta = 0$) and that for these vectors the known relations

$$\frac{d\mathbf{e}_\parallel}{d\varphi} = \mathbf{e}_\perp, \quad \frac{d\mathbf{e}_\perp}{d\varphi} = -\mathbf{e}_\parallel. \qquad (1)$$

hold.[4]

After adding time coordinate $t$ Eqs. (1a) and (1b) lead us to the usual expressions for speed and acceleration in a polar coordinate system.

$$\begin{aligned}\mathbf{r} &= r\,\mathbf{e}_\parallel \\ \mathbf{v} &= \frac{d\mathbf{r}}{dt} = \frac{dr}{dt}\mathbf{e}_\parallel + r\frac{d\varphi}{dt}\mathbf{e}_\perp \\ \mathbf{a} &= \frac{d\mathbf{v}}{dt} = \left(\frac{d^2r}{dt^2} - r\left(\frac{d\varphi}{dt}\right)^2\right)\mathbf{e}_\parallel + \left(r\frac{d^2\varphi}{dt^2} + 2\frac{dr}{dt}\frac{d\varphi}{dt}\right)\mathbf{e}_\perp\end{aligned} \qquad (2)$$

Vectors for momentum and force will be introduced after suggesting a treatment of mass in section 4.

Without going into details, we would like to mention that the issue of possible singularities (e.g. the Schwarzschild singularity leading to black holes) is deeply connected to the treatment of coordinate systems in different theories: In general relativity the Schwarzschild singularity is a property of the coordinate system and can be removed by appropriate coordinate transformations (e.g. to Kruskal coordinates [16]); in alternative theories which are formulated in ordinary spherical coordinates singularities of the Schwarzschild type would pose a physical problem rather than a geometric one, and it is quite natural that such theories rule out their existence.

---

[4] Note that these relations are given in coordinate space which is mathematically flat (i.e. Euclidean). Their validity does not depend on the assumption of a physically meaningful flat "background metric", therefore.

In line with the above clarifications, the notion of the "coordinate perspective" is just a technical one: it summarizes descriptions of distances, speeds, accelerations, energy, mass, momentum, force and relations between these concepts in terms of the coordinates on which the metric is formulated. It does not reveal anything about the underlying nature of the coordinate system itself.

## 2.2 Metric Scheme

The spacetime metrics under consideration are defined by three functions $A(r)$, $B(r)$, $C(r)$ relating coordinate and proper space-time intervals such that the line element for a static, spherically symmetric field becomes

$$ds^2 = A^2 c^2 dt^2 - B^{-2} dr^2 - C^{-2} r^2 \left(d\theta^2 + \sin^2\theta\, d\varphi^2\right). \qquad (3)$$

By this, a time distance of proper length $1$ as seen by the local resting observer is a time distance of length $A^{-1}$ from the coordinate perspective; parallel and perpendicular space distances of proper length $1$ as seen by the local resting observer are parallel and perpendicular distances of lengths $B$ and $C$ from the coordinate perspective. The inverse scaling of time intervals with respect to function $A$ has been chosen such that we have $A=B$ and $C=1$ for the Schwarzschild metric and $A=B=C$ for the exponential metric.

Under the usual assumption that motion takes place in the ($r$, $\varphi$)-plane, the line element (3) reduces to

$$ds^2 = A^2 c^2 dt^2 - B^{-2} dr^2 - C^{-2} r^2 d\varphi^2. \qquad (4)$$

For the analysis of timelike geodesics we use the Langrangian

$$L = \frac{1}{2} g_{\alpha\beta} \frac{dx^\alpha}{d\tau} \frac{dx^\beta}{d\tau} \qquad (5)$$

which leads us to

$$L = \frac{c^2}{2} = \frac{1}{2}\frac{ds^2}{d\tau^2} = \frac{1}{2}\left(A^2 c^2 \left(\frac{dt}{d\tau}\right)^2 - B^{-2}\left(\frac{dr}{d\tau}\right)^2 - C^{-2} r^2 \left(\frac{d\varphi}{d\tau}\right)^2\right). \quad (6)$$

## 2.3 The Geodesic Equations

The Lagrangian (6) represents the first conservation law for geodesic motion. The coordinates $t$ and $\varphi$ are cyclic which allows us to calculate two more conservation laws from the Euler-Lagrange equations (the notations $\dot{t} = dt/d\tau$ and $\dot{\varphi} = d\varphi/d\tau$ being used).

$$p_t = \frac{\partial L}{\partial \dot{t}} = c^2 A^2 \frac{dt}{d\tau} = const \quad (7)$$

$$p_\varphi = \frac{\partial L}{\partial \dot{\varphi}} = -\frac{r^2}{C^2}\frac{d\varphi}{d\tau} = const \quad (8)$$

Exposing the derivatives of the two cyclic variables leads to

$$\frac{dt}{d\tau} = \frac{p_t}{c^2 A^2}, \qquad \frac{d\varphi}{d\tau} = -\frac{p_\varphi C^2}{r^2}. \quad (9)$$

The differentiation of the three conservation laws with respect to $\tau$ leads to the geodesic equations (for derivatives of the functions $A$, $B$ and $C$ with respect to $r$ the prime is used):

Deriving $p_t$ with respect to $\tau$ gives

$$\frac{d^2 t}{d\tau^2} = -2\frac{A'}{A}\frac{dt}{d\tau}\frac{dr}{d\tau}. \quad (10)$$

Deriving $p_\varphi$ with respect to $\tau$ gives

$$\frac{d^2 \varphi}{d\tau^2} = -2\left(\frac{1}{r} - \frac{C'}{C}\right)\frac{dr}{d\tau}\frac{d\varphi}{d\tau}. \quad (11)$$

Deriving $L$ with respect to $\tau$ and inserting from Eqs. (10) and (11) gives

$$\frac{d^2 r}{d\tau^2} = -c^2 A B^2 A'\left(\frac{dt}{d\tau}\right)^2 + \frac{B'}{B}\left(\frac{dr}{d\tau}\right)^2 + r\frac{B^2}{C^2}\left(1 - r\frac{C'}{C}\right)\left(\frac{d\varphi}{d\tau}\right)^2. \quad (12)$$

## 2.4 The Geodesic Equations in *t*-Form

In the following, the coordinate time representation of the geodesic equations will be called the "*t*-form". In order to express the geodesic equations in *t*-form, the variable $\tau$ has to be eliminated. We prepare this by writing:

$$\frac{dt}{d\tau} = \frac{p_t}{c^2 A^2}, \quad \frac{d\varphi}{d\tau} = \frac{d\varphi}{dt}\frac{dt}{d\tau} = \frac{d\varphi}{dt}\frac{p_t}{c^2 A^2},$$
$$\frac{dr}{d\tau} = \frac{dr}{dt}\frac{dt}{d\tau} = \frac{dr}{dt}\frac{p_t}{c^2 A^2} \quad (13)$$

From (7) and (8) we get

$$\frac{d\varphi}{d\tau} = -\frac{p_\varphi C^2}{r^2} = \frac{d\varphi}{dt}\frac{p_t}{c^2 A^2}, \quad p_\varphi = -\frac{p_t r^2}{c^2 A^2 C^2}\frac{d\varphi}{dt}. \quad (14)$$

Differentiating $p_\varphi$ with respect to *t* gives

$$\frac{d^2\varphi}{dt^2} = -2\left(\frac{1}{r} - \frac{A'}{A} - \frac{C'}{C}\right)\frac{dr}{dt}\frac{d\varphi}{dt}. \quad (15)$$

According to (13) *L* can be written as

$$L = \frac{c^2}{2} = \frac{p_t^2}{2c^2 A^2}\left(1 - \frac{1}{c^2 A^2 B^2}\left(\frac{dr}{dt}\right)^2 - \frac{r^2}{c^2 A^2 C^2}\left(\frac{d\varphi}{dt}\right)^2\right) \quad (16)$$

or, using $p_\varphi$ from (14), as

$$L = \frac{c^2}{2} = \frac{1}{2}\left(\frac{p_t^2}{c^2 A^2} - \frac{p_t^2}{c^4 A^4 B^2}\left(\frac{dr}{dt}\right)^2 - \frac{p_\varphi^2 C^2}{r^2}\right). \quad (17)$$

Differentiating with respect to *t* and substituting (7) and (8) for $p_t$ and $p_\varphi$ leads to

$$\frac{d^2 r}{dt^2} = -c^2 A B^2 A' + 2\frac{A'}{A}\left(\frac{dr}{dt}\right)^2 + \frac{B'}{B}\left(\frac{dr}{dt}\right)^2 + r\frac{B^2}{C^2}\left(1 - r\frac{C'}{C}\right)\left(\frac{d\varphi}{dt}\right)^2. \quad (18)$$

The *t*-form of the geodesic equations is given by Eqs. (15) and (18).[5]

---

[5] As a shortcut for deriving Eq. (18), we could also start directly from Eqs. (10) and (12) using the equation $d^2 r/d\tau^2 = (dt/d\tau)^2 d^2 r/dt^2 + (dr/dt)^2 d^2 t/d\tau^2$. This derivation which has been pointed out by an unknown reviewer leaves out the treatment of $p_t$ and $p_\varphi$ which will play a role for the considerations in section 4.1.

## 2.5 Gravitational Acceleration

The expressions for gravitational acceleration **g** in vector notation follow from the relations in (1) and from Eqs. (15) and (18).

$$\mathbf{g} = \mathbf{g}_\parallel + \mathbf{g}_\perp = \left(\frac{d^2 r}{dt^2} - r\left(\frac{d\varphi}{dt}\right)^2\right)\mathbf{e}_\parallel + \left(r\frac{d^2\varphi}{dt^2} + 2\frac{dr}{dt}\frac{d\varphi}{dt}\right)\mathbf{e}_\perp$$

$$\mathbf{g}_\parallel = \left(-c^2 A B^2 A' + 2\frac{A'}{A}\left(\frac{dr}{dt}\right)^2 + \frac{B'}{B}\left(\frac{dr}{dt}\right)^2 + r\left(\frac{B^2}{C^2} - 1 - r\frac{B^2 C'}{C^3}\right)\left(\frac{d\varphi}{dt}\right)^2\right)\mathbf{e}_\parallel \quad (19)$$

$$\mathbf{g}_\perp = 2r\left(\left(\frac{A'}{A} + \frac{C'}{C}\right)\frac{dr}{dt}\frac{d\varphi}{dt}\right)\mathbf{e}_\perp$$

# 3 Applications

Two applications of the geodesic equations, both in original form and in *t*-form, are of special interest in the present context. Inserting the metric coefficients of the Schwarzschild metric leads to the well-known geodesic equations and to the reconstructions of the respective results of McGruder [3] on gravitational repulsion. The geodesic equations of the exponential metric are known, as well. The respective equations in *t*-form will turn out to be the motion equations of the already mentioned SG-model.

## 3.1 The Schwarzschild metric

For the reconstruction of the geodesic equations of the Schwarzschild metric we set:

$$\kappa = \frac{GM}{c^2}, \quad A = B = \left(1 - \frac{2\kappa}{r}\right)^{1/2}, \quad C = 1,$$
$$A' = B' = \frac{\kappa}{r^2}\left(1 - \frac{2\kappa}{r}\right)^{-1/2}, \quad C' = 0 \quad (20)$$

The geodesic equations follow from (10), (11) and (12).

$$\frac{d^2 t}{d\tau^2} = -\frac{2\kappa}{r(r - 2\kappa)}\frac{dt}{d\tau}\frac{dr}{d\tau} \quad (21)$$

$$\frac{d^2 r}{d\tau^2} = -\frac{c^2 \kappa (r - 2\kappa)}{r^3}\left(\frac{dt}{d\tau}\right)^2 + \frac{\kappa}{r(r - 2\kappa)}\left(\frac{dr}{d\tau}\right)^2 + (r - 2\kappa)\left(\frac{d\varphi}{d\tau}\right)^2 \quad (22)$$

$$\frac{d^2\varphi}{d\tau^2} = -\frac{2}{r}\frac{dr}{d\tau}\frac{d\varphi}{d\tau} \qquad (23)$$

The *t*-form equations follow from (18) and (15).

$$\frac{d^2r}{dt^2} = -\frac{c^2\kappa(r-2\kappa)}{r^3} + \frac{3\kappa}{r(r-2\kappa)}\left(\frac{dr}{dt}\right)^2 + (r-2\kappa)\left(\frac{d\varphi}{dt}\right)^2 \qquad (24)$$

$$\frac{d^2\varphi}{dt^2} = -\frac{2(r-3\kappa)}{r(r-2\kappa)}\frac{dr}{dt}\frac{d\varphi}{dt} \qquad (25)$$

This gives for gravitational acceleration

$$\mathbf{g} = \mathbf{g}_\parallel + \mathbf{g}_\perp = \left(-\frac{c^2\kappa(r-2\kappa)}{r^3} + \frac{3\kappa}{r(r-2\kappa)}\left(\frac{dr}{dt}\right)^2 - 2\kappa\left(\frac{d\varphi}{dt}\right)^2\right)\mathbf{e}_\parallel + \frac{2\kappa}{r-2\kappa}\frac{dr}{dt}\frac{d\varphi}{dt}\mathbf{e}_\perp \qquad (26)$$

The parallel component has already been derived by McGruder. The possibility of "gravitational repulsion" follows from the positive term in the parallel component. The condition for positive gravitational acceleration is

$$v_\parallel^2 > \frac{c^2}{3}\left(1-\frac{2\kappa}{r}\right)^2 + \frac{2v_\perp^2}{3}\left(1-\frac{2\kappa}{r}\right), \qquad (27)$$

which is the expression provided by McGruder.

### 3.2 The Exponential Metric

The exponential metric can be treated in the presented metric scheme by setting

$$\kappa = \frac{GM}{c^2}, \quad A = B = C = e^{-\kappa/r} \quad A' = B' = C' = \frac{\kappa}{r^2}e^{-\kappa/r}. \qquad (28)$$

The geodesic equations in original form follow from (10), (11) and (12).

$$\frac{d^2t}{d\tau^2} = -\frac{2\kappa}{r^2}\frac{dt}{d\tau}\frac{dr}{d\tau} \qquad (29)$$

$$\frac{d^2r}{d\tau^2} = -\frac{c^2\kappa}{r^2}e^{-4\kappa/r}\left(\frac{dt}{d\tau}\right)^2 + \frac{\kappa}{r^2}\left(\frac{dr}{d\tau}\right)^2 + (r-\kappa)\left(\frac{d\varphi}{d\tau}\right)^2 \qquad (30)$$

$$\frac{d^2\varphi}{d\tau^2} = -\frac{2(r-\kappa)}{r^2}\frac{dr}{d\tau}\frac{d\varphi}{d\tau} \qquad (31)$$

The *t*-form equations follow from (18) and (15).

$$\frac{d^2r}{dt^2} = -\frac{c^2\kappa}{r^2}e^{-4\kappa/r} + \frac{3\kappa}{r^2}\left(\frac{dr}{dt}\right)^2 + (r-\kappa)\left(\frac{d\varphi}{dt}\right)^2 \quad (32)$$

$$\frac{d^2\varphi}{dt^2} = -\frac{2}{r}\left(1-\frac{2\kappa}{r}\right)\frac{dr}{dt}\frac{d\varphi}{dt} \quad (33)$$

This gives for gravitational acceleration

$$\mathbf{g} = \mathbf{g}_\| + \mathbf{g}_\perp = \left(-\frac{c^2\kappa}{r^2}e^{-4\kappa/r} + \frac{3\kappa}{r^2}\left(\frac{dr}{dt}\right)^2 - \kappa\left(\frac{d\varphi}{dt}\right)^2\right)\mathbf{e}_\| + \frac{4\kappa}{r}\frac{dr}{dt}\frac{d\varphi}{dt}\mathbf{e}_\perp. \quad (34)$$

This is the expression derived by Broekaert. It follows that the motion equations of the SG-model are just the geodesic equations in *t*-form and consequently that free fall in Broekaert's model means geodesic motion in Riemannian spacetime. Broekaert showed that his model explains the four classical tests of GRT (gravitational redshift, light deflection, perihelion precession, and radar echo delay). In the present context this result means yet another support for the respective claim connected with the exponential metric (see footnote 2). The repulsion condition for the exponential metric is

$$v_\|^2 > \frac{1}{3}\left(c^2 e^{-4\kappa/r} + v_\perp^2\right). \quad (35)$$

### 3.3 Two Alternative Formulations of the Schwarzschild Metric

In the present context where the Schwarzschild metric and the exponential metric have been compared it is interesting to take a short look at two reformulations of the Schwarzschild metric.

The Schwarzschild metric in isotropic coordinates[6] is given by setting

$$A = \frac{1-\frac{\kappa}{2r}}{1+\frac{\kappa}{2r}}, \quad B = C = \left(1+\frac{\kappa}{2r}\right)^{-2} \quad (36)$$

and the reformulation suggested by Fock [18] takes the form

---

[6] For the transformation from Schwarzschild coordinates, see e.g. Weinberg [17].

$$A = B = \left(\frac{r-\kappa}{r+\kappa}\right)^{1/2}, \quad C = \left(1+\frac{\kappa}{r}\right)^{-1}. \quad (37)$$

Instead of giving the full (and rather lengthy) expressions for the motion equations we only take a look at the first order approximations which are identical to the respective approximations of the exponential metric, but not to the approximations of the Schwarzschild metric in original form.

$$\frac{d^2 r}{dt^2} = -\frac{c^2 \kappa}{r^2} + \frac{3\kappa}{r^2}\left(\frac{dr}{dt}\right)^2 + (r-\kappa)\left(\frac{d\varphi}{dt}\right)^2 + O\left(\frac{\kappa^2}{r^2}\right) \quad (38)$$

$$\frac{d^2 \varphi}{dt^2} = -\frac{2}{r}\left(1 - \frac{2\kappa}{r}\right)\frac{dr}{dt}\frac{d\varphi}{dt} + O\left(\frac{\kappa^2}{r^2}\right) \quad (39)$$

This result underlines both the similarity of Schwarzschild and exponential metrics and the relevance of coordinate systems.

## 4 Energy and Mass from the Coordinate Perspective

So far, only speeds and accelerations have been described from the coordinate perspective. The following considerations are an attempt to extend the coordinate perspective to energy, mass, momentum and gravitational force. By this, the question whether or not the phenomenon of gravitational repulsion can be regarded as a statement about gravitational force can be answered.

The basic idea for this endeavor is to adapt the special relativistic mass-energy relation $E = mc^2$ such that consistent descriptions of mass and energy of free falling particles become possible. It is inspired by Broekaert's SG-model, but requires further generalization due to the possible dependence of light speeds on orientation with respect to the field, which is not given in the SG-model.

We start with the assumed validity of special relativity for the local resting observer ($u$ stands for the speed of a moving particle as it is measured by the local resting observer; the light speed for the resting observer is $c$ in each direction; $m_0$ is the proper mass and $E_0 = m_0 c^2$ is the proper energy).

$$E_{local} = E_0 \gamma, \quad m_{local} = \frac{E_{local}}{c^2} = \frac{E_0 \gamma}{c^2} = m_0 \gamma, \quad \gamma = \left(1 - \frac{u_\parallel^2}{c^2} - \frac{u_\perp^2}{c^2}\right)^{-1/2} \quad (40)$$

Expressing these statements from the coordinate perspective means to replace the locally measured speed components by the speed components as they are represented in the coordinate perspective:

$$u_\| = \frac{v_\|}{AB}, \quad u_\perp = \frac{v_\perp}{AC}, \quad \gamma = \left(1 - \frac{v_\|^2}{c^2 A^2 B^2} - \frac{v_\perp^2}{c^2 A^2 C^2}\right)^{-1/2} \quad (41)$$

Considering that – from the coordinate perspective - light speed depends on location and orientation relative to the field

$$c_\| = c\,AB, \quad c_\perp = c\,AC \quad (42)$$

this leads to a reformulation of $\gamma$.

$$\gamma_{r,v} \equiv \left(1 - \frac{v_\|^2}{c_\|^2} - \frac{v_\perp^2}{c_\perp^2}\right)^{-1/2} = \gamma = \left(1 - \left(\frac{dr/dt}{c\,AB}\right)^2 - \left(\frac{r\,d\varphi/dt}{c\,AC}\right)^2\right)^{-1/2} \quad (43)$$

For particles resting in the field we define energy in compliance with the treatment of gravitational redshift in general relativity.

$$E(r) \equiv A\,E_0 \quad (44)$$

The different light speeds lead us to introduce two specifications of mass[7] (shortly called "parallel mass" and "perpendicular mass") of a resting particle.

$$m_\|(r) \equiv \frac{E(r)}{c_\|^2}, \quad m_\perp(r) \equiv \frac{E(r)}{c_\perp^2} \quad (45)$$

For the general case of a moving particle we define the energy from the coordinate perspective

$$E \equiv E(r)\gamma_{r,v}, \quad (46)$$

which gives natural definitions for parallel and perpendicular masses of a moving particle.

$$m_\| \equiv m_\|(r)\gamma_{r,v} = \frac{E}{c_\|^2}, \quad m_\perp \equiv m_\perp(r)\gamma_{r,v} = \frac{E}{c_\perp^2} \quad (47)$$

---

[7] There is no connection between the parallel and perpendicular mass concepts introduced here and the concepts of longitudinal and transversal mass used by Einstein [19] in the context of special

## 4.1 Energy, Momentum and Angular Momentum

The definition of energy $E$ of a particle from the coordinate perspective is nothing but a special case of a well-known expression for energy (e.g. Landau & Lifshitz [20]).

$$E = m_0 c^2 \sqrt{g_{00}} \gamma = E_0 A \gamma_{r,v} \qquad (48)$$

This can be seen by starting from Eq. (46) and by the use of Eqs. (44) and (43). While it is known that $E$ is conserved during free fall, if the respective space-time metric is static, there is a simple derivation using the relation to the reformulated Lagrangian (16).

$$E^2 = \frac{E_0^2 p_t^2}{2Lc^2} = \frac{E_0^2 p_t^2}{c^4}, \quad E = \frac{E_0}{c^2} p_t = m_0 p_t = const \qquad (49)$$

The analogue of the special relativistic energy-momentum relation follows from Eq. (16) and from the given definitions.

$$\frac{2LE_0^2}{c^2} = E_0^2 = m_0^2 c^4 = A^{-2} \left( E^2 - c_\|^2 m_\|^2 v_\|^2 - c_\perp^2 m_\perp^2 v_\perp^2 \right)$$
$$= A^{-2} \left( E^2 - c_\|^2 P_\|^2 - c_\perp^2 P_\perp^2 \right) \qquad (50)$$

The respective expression for the exponential metric has already been calculated by Broekaert.

The specification of the two masses gives for the momentum vector **P**

$$\mathbf{P} = P_\| \mathbf{e}_\| + P_\perp \mathbf{e}_\perp \qquad (51)$$

and after inserting from (42) the conservation of angular momentum.

$$P_\varphi \equiv |\mathbf{r} \times \mathbf{P}| = m_\perp r^2 \frac{d\varphi}{dt} = \frac{E}{c_\perp^2} r^2 \frac{d\varphi}{dt} = \frac{E}{c^2 A^2 C^2} r^2 \frac{d\varphi}{dt} = -E \frac{P_\varphi}{p_t} = const \qquad (52)$$

## 4.2 Gravitational force

Using the relations from (1), we get for the parallel and perpendicular force components

$$\mathbf{f} = \frac{d\mathbf{P}}{dt} = \frac{d}{dt}(P_\| \mathbf{e}_\|) + \frac{d}{dt}(P_\perp \mathbf{e}_\perp) = \left( \frac{dP_\|}{dt} - P_\perp \frac{d\varphi}{dt} \right) \mathbf{e}_\| + \left( \frac{dP_\perp}{dt} + P_\| \frac{d\varphi}{dt} \right) \mathbf{e}_\perp = \mathbf{f}_\| + \mathbf{f}_\perp \qquad (53)$$

---

relativity. Einstein's concepts specify the relative orientation of the acceleration of a particle and the speed of the respective particle.

For the derivation we need the time derivatives of the two masses

$$\frac{dm_\parallel}{dt} = E\frac{d}{dt}\left(\frac{1}{c^2 A^2 B^2}\right) = -2m_\parallel\left(\frac{A'}{A} + \frac{B'}{B}\right)\frac{dr}{dt}$$
$$\frac{dm_\perp}{dt} = E\frac{d}{dt}\left(\frac{1}{c^2 A^2 C^2}\right) = -2m_\perp\left(\frac{A'}{A} + \frac{C'}{C}\right)\frac{dr}{dt} \qquad (54)$$

and the relation between parallel and perpendicular masses

$$\frac{m_\perp}{m_\parallel} = \frac{B^2}{C^2} \qquad (55)$$

Insertion from Eqs. (18) and (15) leads us to the two components of the force vector.

$$\mathbf{f}_\parallel = -\left(m_\parallel\left(c_\parallel^2\frac{A'}{A} + v_\parallel^2\frac{B'}{B}\right) + m_\perp v_\perp^2\frac{C'}{C}\right)\mathbf{e}_\parallel$$
$$= -E\left(\frac{A'}{A} + \frac{B'}{B}\frac{v_\parallel^2}{c_\parallel^2} + \frac{C'}{C}\frac{v_\perp^2}{c_\perp^2}\right)\mathbf{e}_\parallel \qquad (56)$$
$$\mathbf{f}_\perp = (m_\parallel - m_\perp)\frac{v_\parallel v_\perp}{r}\mathbf{e}_\perp = \frac{E}{c^2 A^2}\left(\frac{1}{B^2} - \frac{1}{C^2}\right)\frac{v_\parallel v_\perp}{r}\mathbf{e}_\perp$$

The perpendicular force vector vanishes only for $B=C$. In the context of general relativity, this makes the already mentioned isotropic representation and Focks's representation of the Schwarzschild solution more appealing than the original formulation. In the context of alternative models built on a Euclidean background metric this might be interpreted as support of the assumption of "an isotropic gravitational effect on lengths".

In the Schwarzschild case where $B=A$ and $C=1$ we get from (56)

$$\mathbf{f}_\parallel = -m_\parallel\frac{A'}{A}(c_\parallel^2 + v_\parallel^2)\mathbf{e}_\parallel = -E\frac{A'}{A}\left(1 + \frac{v_\parallel^2}{c_\parallel^2}\right)\mathbf{e}_\parallel$$
$$\mathbf{f}_\perp = (m_\parallel - m_\perp)\frac{v_\parallel v_\perp}{r}\mathbf{e}_\perp = \frac{E}{c^2 A^2}\left(\frac{1}{A^2} - 1\right)\frac{v_\parallel v_\perp}{r}\mathbf{e}_\perp \qquad (57)$$

which shows an independence of the parallel force component on the perpendicular speed component, while there does exist a dependence on the perpendicular speed component for gravitational acceleration (Eq. (19)).

For the exponential metric we have $A=B=C$ and consequently no dependence of light speed and mass on orientation. This gives for gravitational force

$$\mathbf{f} = \mathbf{f}_\parallel = -m\frac{A'}{A}\left(c(r)^2 + v_\parallel^2 + v_\perp^2\right)\mathbf{e}_\parallel$$
$$= -E\frac{A'}{A}\left(1 + \frac{v^2}{c(r)^2}\right)\mathbf{e}_\parallel = -E\frac{A'}{A}(2 - \gamma_{v,r}^{-2})\mathbf{e}_\parallel \qquad (58)$$

which is the force expression of the SG-model.

The main result of the present considerations on gravitational force is the impossibility of a positive parallel force component for both the Schwarzschild and the exponential metric which can be read from Eqs. (57) and (58). The suggested treatment of the mass-energy relation thus makes gravitational repulsion a phenomenon that concerns only gravitational acceleration.

## 5 Conclusions

The geodesic equations for the general class of diagonal metrics of spherically symmetric fields have been calculated, both in the original form with respect to proper time and in the "$t$-form" with respect to coordinate time. Inserting the metric coefficients for the Schwarzschild metric has led to the known geodesic equations and to the results of McGruder on "gravitational repulsion". In the case of the exponential metric the derived $t$-form equations are identical with the motion equations of a "scalar gravity model", which has been formulated by Broekaert without the use of pseudo-Riemannian geometry.

The adaptation of the special relativistic mass-energy relation to local and possibly anisotropic light speeds which is motivated by Broekaert's approach has led to conservation laws for energy and angular momentum and to an account of gravitational force.

It turned out that from the suggested viewpoint there is no repulsive gravitational force, neither for the Schwarzschild metric nor for the exponential metric.

We would finally like to note that the present work is of technical nature and does not provide any arguments for or against alternative approaches to gravitation. The purpose has rather been to give an account of the coordinate perspective of geodesic motion for the class of static, spherically symmetric spacetimes that is as general as possible and that does not depend on the theory that stands behind a specific spacetime metric. Whether or not the introduced concepts are valuable for the debate about alternative gravitation theories remains to be seen.